\documentclass
[preprint,prl,twocolumn,tightenlines,10pt,notitlepage,superscriptaddress]{revtex4}%
\usepackage{amssymb}
\usepackage{amsfonts}
\usepackage{amsmath}
\usepackage{graphicx}%
\providecommand{\U}[1]{\protect\rule{.1in}{.1in}}

\begin{document}
\preprint{ }
\title[Number theory and quantum spin systems]{Number-Theoretic Nature of Communication in Quantum Spin Systems}
\author{Chris Godsil}
\affiliation{Combinatorics \& Optimization, University of Waterloo, N2L 3G1 Waterloo, Canada}
\author{Stephen Kirkland}
\affiliation{Hamilton Institute, National University of Ireland, Maynooth, County Kildare, Ireland}
\author{Simone Severini}
\affiliation{Department of Computer Science, and Department of Physics \& Astronomy,
University College London, WC1E 6BT London, United Kingdom}
\author{Jamie Smith}
\affiliation{Institute for Quantum Computing, and Combinatorics \& Optimization, University
of Waterloo, N2L 3G1 Waterloo, Canada}
\keywords{Spin chains; state transfer; prime numbers}
\pacs{03.67.Hk, 05.50.+q, 32.80.Lg}

\begin{abstract}
The last decade has witnessed substantial interest in protocols for
transferring information on networks of quantum mechanical objects. A variety
of control methods and network topologies have been proposed, on the basis
that transfer with perfect fidelity --- \emph{i.e.} deterministic and without
information loss --- is impossible through unmodulated spin chains with more
than a few particles. Solving the original problem formulated by Bose
[\emph{Phys. Rev. Lett.} \textbf{91}, 207901 (2003)], we determine the exact
number of qubits in unmodulated chains (with XY\ Hamiltonian) that permit the
transfer with fidelity arbitrarily close to $1$, a phenomenon called
\emph{pretty good state transfer}. We prove that this happens if and only if
the number of nodes is $n=p-1$, $2p-1$, where $p$ is a prime, or $n=2^{m}-1$.
The result highlights the potential of quantum spin system dynamics for
reinterpreting questions about the arithmetic structure of integers, and, in
this case, primality.

\end{abstract}
\volumeyear{ }
\volumenumber{ }
\issuenumber{ }
\eid{ }
\date{24 June 2012}
\startpage{1}
\endpage{6}
\maketitle

\emph{Introduction. --- }Since the pioneering work by Bose \cite{bo0}, quantum
channels implemented by spin systems have been interpreted as wires\ for the
transmission of possibly unknown qubit states. Motivated by the perspective of
designing quantum buses and nanodevices, a vigorous effort has delineated the
field of quantum spin systems engineering \cite{bo}. In this context, one of
the desirable tasks is to transfer the state of a particle into another one
with maximum fidelity; when the fidelity is $1$, we have \emph{perfect state
transfer }(\emph{PST}), a notion originally introduced in \cite{c}. Given an
inherent difficulty in the manipulation of coherent quantum mechanical
objects, the most appealing set up for state transfer employs a
time-independent Hamiltonian and no interaction with the system except at
initialisation and read-out. If we do not use \emph{ad hoc} coupling schemes
and control protocols (see \cite{kay}; also the recent \cite{ch} and the
references therein), a spin chain (1D magnet), with a Heisenberg $XY$
Hamiltonian, exhibits PST only between 2 and 3 qubits \cite{c}. This is a
negative result because such a chain is arguably the quantum wire obtainable
with the smallest amount of physical and technological resources.

On the basis of sufficient conditions for PST, this fact fueled a
mathematically challenging classification programme aimed to identify PST in
general network topologies \cite{go}. The $k$-dimensional hypercube (with
$n=2^{k}$ qubits) has the best known performance in terms of the distance
travelled by a single excitation: PST occurs between antipodal nodes at
(network)$\ $distance $k$. However, for practical purposes, it is natural to
study whether a minimalist structure like the chain (with $n$ qubits) can
still be usefully employed. In particular, even if we already know that there
is no PST for $n\geq4$ we may still ask the following question: given $n$ and
an $\varepsilon>0$, does there exist $t$ such that the fidelity at time $t$
between qubits $1$ and $n$ is larger than $1-\varepsilon$? When the answer is
\textquotedblleft yes\textquotedblright, we say that there is \emph{pretty
good state transfer} (\emph{PGST}). While Bose \cite{bo0} verified that the
fidelity could be remarkably high even for rather long chains, the notion of
PGST was formally isolated in \cite{god1} as a relaxation of PST. We give, in
the present Letter, a complete characterization of the parameters for which
there is PGST. Our findings can be combined into the following simple-sounding statement:

\noindent\emph{Theorem.} -- A uniformly coupled chain of $n$ particles with
$XY$ Hamiltonian has PGST if and only if $n=p-1$ or $2p-1$, where $p$ is a
prime, or if $n=2^{m}-1$.

The significance of the result is twofold. From the physical point of view, it
is valuable that we rigorously describe a phenomenon with applications to the
study of quantum nanodevices \cite{di2000}. In fact, the related observations
obtained up to now are either numerical or fragmented. The message conveyed by
the notion of PGST is that an unknown qubit state can be transferred with
arbitrarily large fidelity between the end nodes of \textquotedblleft
long\textquotedblright\ chains, whenever the waiting time is not an issue.
From the mathematical point of view, we highlight properties of quantum
interference as a consequence of number-theoretic constraints. The appearance
of prime numbers in the theorem indicates a connection between the occurrence
of PGST in chains of a given length and primality testing. This suggests a
potential application of quantum dynamics on graphs for reinterpreting
number-theoretic problems, when the problem description is encoded in the
parameters of the system.

It is well established that the dynamics of a single excitation in networks of
spins with unmodulated couplings is a continuous-time quantum walk on the
unweighted graph modeling the network. Hence, our result can also be
reinterpreted in the language of continuous-time quantum walks: it describes
the maximum hitting probability between antipodal vertices induced by a
one-dimensional quantum walk. Walks of this type have been studied in great
detail \cite{p03}. This analogy prompts us to disregard the original spin
system setup and to work with a single $n$-level system. The realisation and
simulation of this quantum device is the centre of several discussions
\cite{sc09}. It is remarkable that at a wider level, state transfer can be
seen as the simplest model in a family of processes for quantum transport: for
example, the population transfer of $n$-level systems \cite{sh2008}, after
including the role of the environment, the transport of energy in organic
molecules \cite{re08}, and a closely related mathematical setting describes
state transfer in chains of harmonic oscillators coupled with beam-splitter
interactions \cite{pl04}. Finally, the experimental implementation of the
Heisenberg $XY$ chain has been proposed in schemes including cold-atom optical
lattices and superconducting circuits\emph{ }\cite{m}.

\emph{PGST. -- }The Hamiltonian governing the evolution of the system acts on
the Hilbert space $\mathcal{H}\cong\mathbb{C}_{1}^{2}\otimes\cdots
\otimes\mathbb{C}_{n}^{2}$. If we do not include external static potentials,
the Hamiltonian is
\[
\widehat{H}=\frac{1}{2}\sum_{u=1}^{n-1}J_{u}\left(  \sigma_{u}^{x}\sigma
_{u+1}^{x}+\sigma_{u}^{y}\sigma_{u+1}^{y}\right)  ,
\]
where $\sigma_{u}^{w}$ ($w\in\{x,y,z\}$) is a Pauli matrix on $\mathbb{C}%
_{u}^{2}$ and $J_{u}$ is the coupling strength between the particles $u$ and
$u+1$. By virtue of the Jordan-Wigner transform \cite{l}, the free evolution
for a time $t$ of a single excitation originally located at site $|u\rangle
\in\{|1\rangle,...,|n\rangle\}$ is given by $e^{iHt}|u\rangle=U(t)|u\rangle$,
where $H_{u,v}=J_{u}$ if $v=u+1$ or $u=v+1$, and $H_{u,v}=0$, otherwise. The
$n\times n$ real symmetric matrix $H$ is the Hamiltonian restricted to the
single excitation sector. PST occurs between $1$ and $n$ if there is a
$t\in\mathbb{R}^{+}$ such that $|\langle n|U(t)|1\rangle|=1$, \emph{i.e.}, the
channel has maximum fidelity, PGST occurs between $1$ and $n$ if for every
$\epsilon>0$ there is $t\in\mathbb{R}^{+}$ such that $|\langle n|U(t)|1\rangle
|>1-\epsilon$. Analytical solutions for coupling design able to achieve PST
for any $n$ have been presented in previous works (see \cite{ch}). Abstractly,
a chain of length $n$ is modeled by a network called an $n$\emph{-path} and
denoted by $P_{n}$. The links representing the particle-particle interactions
are $\{1,2\},\{2,3\},...,\{n-1,n\}$. When the chosen couplings are uniform
(\emph{w.l.o.g.}, $J_{u}=1$), the Hamiltonian $H$ is the adjacency matrix of
the $n$-path. (Recall that the adjacency matrix of a graph has $ij$-th entry
$1$ if there is a link between the nodes $i$ and $j$; $0$, otherwise.) We know
from \cite{god1} that there is PGST in $P_{4}$ and $P_{5}$.

We shall first prove the theorem. Next, we present some details about the
cases when there is no PGST. In particular, we will give an explicit upper
bound on the fidelity in a special case. In Appendix 3, we discuss PGST
between internal nodes, by considering a link between state transfer and
control theory on networks (see \cite{bu}).

\emph{Proof of the theorem. ---} The proof of the theorem is based on a direct
linear-algebraic analysis of the eigensystem of $U(t)$ and on the application
of standard number-theoretic tools, especially Kronecker's theorem on
Diophantine approximation. We begin by considering a general property of PGST
and basic facts about bipartite graphs. We then use relations on the
eigenvalues with a detailed case-by-case treatment.

For any two vertices $u$ and $v$ of a graph,
\[
U(t)|u\rangle-\gamma|v\rangle=\gamma U(t)(\gamma^{-1}|u\rangle-U(-t)|v\rangle
);
\]
here
$\gamma^{-1}I-U(-t)$ is the Hermitian adjoint of $\gamma^{-1}I-U(t)$. Since
$U(t)$ is unitary and $\Vert\gamma\Vert=1$, we see that
\[
\Vert U(t)|u\rangle-\gamma|v\rangle\Vert=\Vert U(t)|v\rangle-\gamma
|u\rangle\Vert.
\]
So, if we have PGST from $u$ to $v$, we also have it from $v$ to $u$.

A graph $X$ is \emph{bipartite} if there is a bipartition of the set of nodes
such that the links connect only nodes in different parts. Suppose $X$ is
bipartite and let $D$ be a diagonal matrix such that $D_{u,u}$ is $1$ or $-1$,
accordingy as $u$ is in one or the other part of the bipartition. Then
$DAD=-A$, and if $U(t)|u\rangle\approx\gamma|v\rangle$,
\[
\gamma D|v\rangle\approx DU(t)D\,D|u\rangle=U(-t)D|u\rangle.
\]
But $|u\rangle$ and $|v\rangle$ are eigenvectors for $D$ with eigenvalues $1$
or $-1$; the eigenvalues are equal if and only if $u$ and $v$ are in the same
part. So there is a sign factor $\sigma_{u,v}$ and $\gamma|v\rangle
\approx\sigma_{u,v}U(-t)|u\rangle$. Accordingly, $U(t)|v\rangle\approx
\gamma^{-1}\sigma_{u,v}|u\rangle$.

By the above, $U(t)|v\rangle\approx\gamma|v\rangle$, and we conclude that
$\gamma\approx\gamma^{-1}\sigma_{u,v}$. Hence, $\gamma\approx\pm1$ if $u$ and
$v$ are in the same part, and $\gamma\approx\pm i$ if they are not. (For PST
this observation is due to Kay \cite{bo}.) Let $F$ denote the permutation
matrix of order $n\times n$ such that $F|r\rangle=|n+1-r\rangle$ for all $r$.
Let $E_{1},\ldots,E_{n}$ be the idempotents in the spectral decomposition of
the path $P_{n}$ (\emph{i.e.}, of its adjacency matrix). We can then write
$F=\sum_{r=1}^{n}(-1)^{r-1}E_{r}$. If we have PGST at time $t$ then
$U(t)\approx\gamma F$ and, therefore,
\[
1=\det U(t)\approx\gamma^{n}\det(F)=\gamma^{n}(-1)^{\lfloor n/2\rfloor}.
\]
This yields three cases: \emph{(1)}$\ $if $n\equiv1$ mod 4 then $(-1)^{\lfloor
n/2\rfloor}=1$ and $\gamma\approx1$; \emph{(2)}$\ $if $n\equiv3$ mod 4 then
$(-1)^{\lfloor n/2\rfloor}=-1$ and $\gamma\approx-1$; \emph{(3)}$\ $if $n$ is
even then $i^{n}=(-1)^{n/2}$ and $\gamma\approx\pm i$. It is well known that
the eigenvalues $\theta_{r}$ of $P_{n}$ are given by $\theta_{r}=2\cos[\pi
r/(n+1)]$.

We start with the positive results. If PGST occurs then $U(t)$ gets
arbitrarily close to $\gamma F$. This means that $e^{i\theta_{r}t}%
\approx(-1)^{r-1}\gamma$ for $r=1,\ldots,n$. Set $m=\lfloor n/2\rfloor$.
Assume $\gamma=\pm1$ if $n$ is odd and $\pm i$ if $n$ is even.

First, we prove that for the path $P_{n}$, if $e^{i\theta_{r}t}\approx
(-1)^{r-1}\gamma$, for $r=1,\ldots,m$, then $e^{i\theta_{r}t}\approx
(-1)^{r-1}\gamma$, for all $r=1,...,n$, and, hence, $U(t)\approx\gamma F$. To
see this, assume $e^{i\theta_{r}t}\approx(-1)^{r-1}\gamma$. Since $n$-paths
are bipartite, $\theta_{n+1-r}=\theta_{r}$, and, therefore, $e^{i\theta
_{n+1-r}t}=e^{-i\theta_{r}t}\approx(-1)^{r-1}\gamma^{-1}$. For PGST, we need
$(-1)^{n-r}\gamma=(-1)^{r-1}\gamma^{-1}$, or, equivalently, $\gamma
^{2}=(-1)^{n-1}$. As this holds for our choice of $\gamma$, we are done. Since
$\theta_{r}=2\cos[\pi r/(n+1)]$ we have that $\theta_{r}$ is a polynomial with
rational coefficients of degree $r$ in $\theta_{1}$. We also set $\theta
_{0}=2$. It follows that the numbers $\theta_{0},\ldots,\theta_{d}$ are
linearly independent over $\mathbb{Q}$ if and only if the degree of the
algebraic integer $\theta_{1}$ is greater than $d$.

This provides us with the necessary tools to prove the first part of the
theorem: if $n=p-1$ or $2p-1$, where $p$ is prime, or if $n=2^{m}-1$ then we
have PGST on $P_{n}$. If $\alpha\in\mathbb{C}$ we use use $\mathbb{Q}(\alpha)$
to denote the field obtained by adjoining $\alpha$ to $\mathbb{Q}$. If
$\mathbb{F}$ is a subfield of $\mathbb{Q}(\alpha)$, then $\mathbb{Q}(\alpha)$
is a vector space over $\mathbb{F}$; its dimension is the index of
$\mathbb{F}$ in $\mathbb{Q}(\alpha)$. (For details see, \emph{e.g.},
\cite{her}, Chap. 5.) Let $\theta=2\cos[\pi/(n+1)]$ and set $\zeta
=e^{i\pi/(n+1)}$. Then $\theta\in\mathbb{Q}(\zeta)$ and $\zeta$ is a root of
the quadratic $x^{2}-x\theta+1$. So the index of $\mathbb{Q}(\theta)$ in
$\mathbb{Q}(\zeta)$ is at most two. If $n\geq3$ though, $\theta$ is real and
$\zeta$ is not. Thus, the index is exactly two. The degree of $\zeta$ is
$\phi(2n+2)$, where $\phi$ is Euler's function, and therefore, the degree of
$\theta$ is $\phi(2n+2)/2$. If $n=p-1$, $2p-1$, or $2^{m}-1$, where $p$ is a
prime, we have, respectively
\begin{align*}
\phi(2n+2) &  =\phi(2p)=\phi(p)=p-1,\\
\phi(2n+2)/2 &  =\phi(4p)/2=\phi(p)=p-1,\\
\phi\lbrack2(n+1)]/2 &  =2^{m-1}.
\end{align*}
Finally, in each of these three cases the positive eigenvalues of $P_{n}$ are
linearly independent over $\mathbb{Q}$. Next, let $\delta$ equal $0,\frac
{1}{2},\frac{1}{4},$ or $\frac{3}{4}$, accordingly $\gamma$ equals
$1,-1,i,-i,$ respectively. For each $r=1,\ldots,m,$ let $\alpha_{r}=\frac
{1}{2}$ if $r$ is even and $\alpha_{r}=0$ if $r$ is odd. By Kronecker's
theorem (see \cite{hw}), for any $\epsilon,T>0$ there is a $t_{\epsilon}>T$,
and integers $p_{r},r=1,\ldots,m,$ such that $\left\vert t_{\epsilon}%
\frac{\theta_{r}}{2}-p_{r}-\alpha_{r}-\delta\right\vert <\epsilon$. It follows
that as $\epsilon\rightarrow0$, $e^{i\pi t_{\epsilon}\theta_{r}}%
\rightarrow(-1)^{r-1}\gamma$, with $r=1,\ldots,n$, so that we have PGST. Let
us now focus on the negative cases: if $n=mp-1$, where $p$ is odd and $m\geq
3$, then PGST does not occur on $P_{n}$. Suppose $n+1=mp$, where $p$ is odd.
After some algebra (see Appendix 1), we obtain the following equation:%
\begin{multline*}
(\theta_{1}-\theta_{2})+%
{\displaystyle\sum\limits_{r=1}^{\frac{p-1}{2}}}
(-1)^{r}(\theta_{mr+1}-\theta_{mr+2})\\
+%
{\displaystyle\sum\limits_{r=1}^{\frac{p-1}{2}}}
(-1)^{r}(\theta_{mr-1}-\theta_{mr-2})=0.
\end{multline*}
Denote the three terms on the left by $D$, $E$, and $F$, respectively. If we
have PGST, then there is a sequence of times $(t_{k})_{k\geq0}$ such that
$e^{i\theta_{r}t_{k}}\rightarrow(-1)^{r-1}\gamma$, and so $e^{i(\theta
_{s}-\theta_{s+1})t_{k}}\rightarrow-1$. Therefore, $e^{iDt_{k}}\rightarrow-1$,
while $e^{iEt_{k}}$ and $e^{iEt_{k}}$ both tend to $1$ or to $-1$. Thus,
$e^{i(D+E+F)t_{k}}\rightarrow-1$, which is impossible, since $D+E+F=0$. It is
not difficult to verify that the cases considered above include all integers.
This ends the proof of the theorem. A corollary is that if $2\leq n\leq10$, we
have always PGST, except in the case $n=8$.%

\begin{figure}
[h]
\begin{center}
\includegraphics[
height=1.2618in,
width=1.8375in
]%
{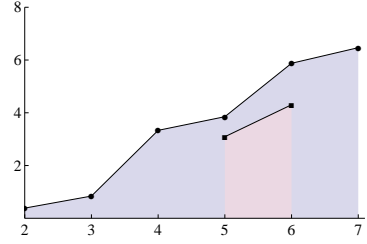}%
\caption{Logarithm of the earliest points in which the fidelity is strictly
greater than $0.99$ for $2\leq n\leq7$. The numbers have been obtained by
plotting $|U(t)_{1,n}|$ ($n=2,...,7$) and then by analysing sections of the
curves. Clearly $0.99$ is an arbitrary choice. Notice the jumps between the
pairs $(2,3),(4,5),(6,7)$. Because PGST depends on the positive eigenvalues of
$P_{n}$, this phenomenon may be explained by the fact that $\left\lfloor
n/2\right\rfloor =\left\lfloor (n+1)/2\right\rfloor $, for $n$ even. }%
\label{pgst17}%
\end{center}
\end{figure}
Fig. \ref{pgst17} gives the smallest times needed to achieve a relatively
large fidelity ($>0.99$ and, thus, $\epsilon=0.01$) for chains of length
$2\leq n\leq7$. The parameters to be considered for numerics are $n$ and
$\epsilon$. Indeed, the waiting time depends also on the tolerance
$\varepsilon$ being close to one. The log plot suggests that for a fixed
$\varepsilon$ there is a behaviour that is linear in $n$.

It is a corollary of the theorem that there is PGST on $P_{n}$ if and only if
its positive eigenvalues are linearly independent over the rationals. A proof
of some cases when there is PGST can be constructed with the use of facts
about linear independence of roots of unity discussed by Conway and Jones (in
particular Theorems 1 and 7) and by Watkins and Zeitlin \cite{cj}. The crucial
observation for PGST is indeed the linear independence (over the rationals) of
the numbers $\cos[\pi j/n+1]$ for certain choices of $j$.

\emph{Bounding fidelity. ---} We have seen that there are many cases in which
there is no PGST for $P_{n}$. We outline now a general technique for proving
upper bounds on the fidelity whenever this happens. Algebraic graph theory is
again the natural toolbox to employ. The spectral decomposition of the
adjacency matrix of a graph $X$ is $A=\sum_{r}\theta_{r}E_{r}$. Two vertices
$u$ and $v$ of $X$ are \emph{cospectral} if, for each $r$, the projections
$E_{r}|u\rangle$ and $E_{r}|v\rangle$ have the same length. We say they are
\emph{strongly cospectral} if, for each $r$, we have $E_{r}|u\rangle=\pm
E_{r}|v\rangle$. In \cite{god1}, it is shown that if we have PGST (or PST)
from $u$ to $v$ then $u$ and $v$ are strongly cospectral. If the eigenvalues
of $A$ are simple, two vertices are strongly cospectral if and only if they
are cospectral. Assume $U(t)_{u,v}=\sum_{r}(E_{r})_{u,v}e^{i\theta_{r}t}$.
Define $\epsilon_{r}$ by the requirement that $(E_{r})_{u,v}=\epsilon
_{r}(E_{r})_{u,u}$. (For paths, $\epsilon_{r}=(-1)^{r-1}$.) Then $U(t)_{u,v}$
is a convex combination of the norm one complex numbers $\epsilon
_{r}e^{it\theta_{r}}$. For PGST to occur, these numbers must all be
approximately equal. We can see this by applying the triangle inequality,
\[
|U(t)_{u,v}|\leq\sum_{r}|(E_{r})_{u,v}|=\sum_{r}|(E_{r})_{u,u}|=1.
\]
In particular, PGST cannot happen if there is some set $S$ of eigenvalue
indices such that
\[
\sum_{r\in S}(E_{r})_{u,u}-\left\vert \sum_{r\in S}(E_{r})_{u,v}e^{i\theta
_{r}t}\right\vert
\]
is bounded away from zero (for all $t$). With a direct analysis of this
expression, we can rule out PGST explicitly when $n=3k+2$ and $k$ is even.
(The details are in Appendix 2.) We leave open the challenge of finding
explicit bounds on the fidelity in the remaining cases. The proof technique
outlined here can be potentially extended to other network topologies.

\emph{Conclusions. ---} By solving an open problem about quantum transport
\cite{bo0}, we have highlighted number-theoretic properties of quantum
communication in spin chains. We have studied general properties of PGST. We
have given necessary and sufficient mathematical conditions for PGST to occur
on $XY$ spin chains with uniform couplings.

The physical intuition paralleling the mathematical result suggests that the
\emph{spin wave} can reach the end of the chain with an arbitrary high peak
only when the number of particles does not permit significant constructive
interference. In this case, the trajectory of the amplitude (in a fixed time
interval) suggests an intriguing analogy with chaotic dynamics that remains to
be explored.

Deciding whether there is PGST is computationally equivalent to primality
testing, a task that is performed efficiently with the AKS test \cite{aks}.
Experimental detection of PGST would correspond to a \emph{natural algorithm}
for primality. Its complexity would be determined via bounds on the time
required by physical evolution (or its simulation) and on resources for
tomography and sequential measurements.

We have outlined a technique for quantifying maximum fidelity when there is no
PGST. Some cases remain with no complete answer. We have shown how notions of
network control theory can be applied to study communication in spin systems.
Exploring PGST in general networks beyond the $n$-path requires similar
methods, but it is a challenging task. It is valuable to observe that we have
considered a system without spatial disorder; its behaviour does not exhibit
effects due to Anderson localization. Numerics in \cite{bu0} indicated that
the fidelity of this system tends to be robust when a relatively small amount
of disorder is introduced in the couplings. On the other side, it was shown in
\cite{kea}\emph{ }that the speed of propagation of coherent walks is
suppressed exponentially in the amount of imperfection.

We have left open the development of a comprehensive theory of PGST. Such a
theory is important to obtain a fuller understanding of transport in networks
of quantum mechanical particles, either engineered or found in nature.
Experimental tests based on photonic waveguides are currently investigated.
(See \cite{per} for background on such schemes.)

\noindent\emph{Acknowledgments. --- }We thank Sougato Bose, Xiaoxia Fan,
Alastair Kay, Avinash Kolli, Mike Pepper, Tommaso Tufarelli, and Sanju Velani
for useful conversation and interest in this work. We thank the anonymous
referees for their comments and suggestions. C. G. and J. S. acknowledge
support from NSERC; S. K. from the Science Foundation Ireland (under Grant No.
SFI/07/SK/I1216b); S. S. from the Royal Society.

\newpage

\emph{Appendix 1 -- }In the proof of the theorem, suppose $n+1=mp$, where $p$
is odd. Then
\[
1+2\sum_{r=1}^{(p-1)/2}(-1)^{r}\cos\left(  \pi r/p\right)  =0.
\]
If we multiply this by $\cos\left(  \pi/\left(  n+1\right)  \right)  $, we
get
\begin{multline*}
\cos\left(  \frac{\pi}{n+1}\right)  +\sum_{r=1}^{\frac{p-1}{2}}(-1)^{r}\\
\left[  \cos\left(  \frac{\pi(mr+1)}{n+1}\right)  +\cos\left(  \frac
{\pi(mr-1)}{n+1}\right)  \right]  =0,
\end{multline*}
which yields the following relations on eigenvalues:
\[
\theta_{x}+\sum_{r=1}^{\frac{p-1}{2}}(-1)^{r}\theta_{mr+x}+\sum_{r=1}%
^{\frac{p-1}{2}}(-1)^{r}\theta_{mr-x}=0,
\]
with $x=1,2$. The equation with $x=2$ is obtained if we multiply by
$\cos\left(  2\pi/(n+1)\right)  $ the equation with $x=1$. If we subtract
these equations, we have
\begin{multline*}
(\theta_{1}-\theta_{2})+%
{\displaystyle\sum\limits_{r=1}^{\frac{p-1}{2}}}
(-1)^{r}(\theta_{mr+1}-\theta_{mr+2})\\
+%
{\displaystyle\sum\limits_{r=1}^{\frac{p-1}{2}}}
(-1)^{r}(\theta_{mr-1}-\theta_{mr-2})=0.
\end{multline*}

\bigskip

\emph{Appendix 2 -- }We have outlined a general technique for bounding
fidelity. We know that there is no PGST when $n=3k+2$ and $k$ is even. This is
the case considered here. There is no PGST if there exists some set $S$ of
eigenvalue indices such that the expression
\[
\sum_{r\in S}(E_{r})_{u,u}-\left\vert \sum_{r\in S}(E_{r})_{u,v}e^{i\theta
_{r}t}\right\vert
\]
is bounded away from zero (for all $t$). In this case, $\theta_{1}=\theta
_{k}+\theta_{k+2}$. Define
\[
h(t)=\sum_{r}(E_{r})_{1,n}e^{i\theta_{r}t}.
\]
Here,
\[
(E_{r})_{1,n}=(-1)^{r-1}(E_{r})_{1,1},
\]
where $(E_{r})_{1,1}\geq0$ and $\sum_{r}(E_{r})_{1,1}=1$. So, we can rewrite
$h(t)$ in the form%
\[
h(t)=\sum_{r}a_{r}(-1)^{r-1}e^{i\theta_{r}t}.
\]
Since $\sum_{r}a_{r}=1$, if $|h(t)|\approx1$ then the summands in this
expression must be approximately equal. Consider the sum
\begin{align*}
&  a_{1}e^{i\theta_{1}t}+(-1)^{k-1}a_{k}e^{i\theta_{k}t}\\
&  +(-1)^{k+1}a_{k+2}e^{i\theta_{k+2}t}+(-1)^{n}a_{n}e^{i\theta_{n}t}.
\end{align*}
For PGST to occur, its absolute value must be close to $a_{1}+a_{k}%
+a_{k+2}+a_{n}$. We can simplify a little by working with
\begin{align*}
&  a_{1}+(-1)^{k-1}a_{k}e^{i(\theta_{k}-\theta_{1})t}\\
&  +(-1)^{k+1}a_{k+2}e^{i(\theta_{k+2}-\theta_{1})t}+(-1)^{n}a_{n}%
e^{i(\theta_{n}-\theta_{1})t};
\end{align*}
which has the same absolute value. We aim to show that the real part of this
sum is bounded away from $a_{1}+a_{k}+a_{k+2}+a_{n}$. We note that $\theta
_{n}=-\theta_{1}$, $a_{n}=a_{1}$, and each of $a_{1},a_{k},a_{k+2},a_{n}$ is positive.

When $k$ is even, we can write the real part of this sum as
\begin{align*}
&  a_{1}-a_{k}\cos[(\theta_{k}-\theta_{1})t]\\
&  -a_{k+2}\cos[(\theta_{k+2}-\theta_{1})t]-a_{1}\cos(2\theta_{1}t).
\end{align*}
Since $\theta_{1}=\theta_{k}+\theta_{k+2}$, this is equal to
\begin{align*}
&  a_{1}-a_{k}\cos(\theta_{k+2}t)\\
&  -a_{k+2}\cos(\theta_{k}t)-a_{1}\cos[2(\theta_{k}+\theta_{k+2})t].
\end{align*}
When
\[
\cos(\theta_{k+2}t),\cos(\theta_{k}t)\leq-\sqrt{3}/2,
\]
we have%
\[
-1/2\leq\sin(\theta_{k+2}t),\sin(\theta_{k}t)\leq1/2.
\]
Whence
\[
\cos[(\theta_{k}+\theta_{k+2})t]\geq\left(  3/4-1/4\right)  =1/2
\]
and so%
\[
\cos[2(\theta_{k}+\theta_{k+2})t]\geq-1/2.
\]
Consequently,
\begin{align*}
&  a_{1}-a_{k}\cos(\theta_{k+2}t)-a_{k+2}\cos(\theta_{k}t)\\
&  -a_{1}\cos[2(\theta_{k}+\theta_{k+2})t]\\
&  \leq3a_{1}/2+(a_{k}+a_{k+2})\sqrt{3}/2\\
&  =2a_{1}+a_{k}+a_{k+2}-(a_{1}+(2-\sqrt{3})a_{k}\\
&  +(2-\sqrt{3})a_{k+2})/2.
\end{align*}
On the other hand, if $\cos(\theta_{k+2}t)\geq\sqrt{3}/2$ then
\begin{align*}
&  a_{1}-a_{k}\cos(\theta_{k+2}t)-a_{k+2}\cos(\theta_{k}t)\\
&  -a_{1}\cos[2(\theta_{k}+\theta_{k+2})t]\\
&  \leq2a_{1}+\sqrt{3}a_{k}/2+a_{k+2}+a_{1}\\
&  =2a_{1}+a_{k}+a_{k+2}\\
&  -(2-\sqrt{3})a_{k}/2.
\end{align*}
It follows that if $\cos(\theta_{k}t)\geq\sqrt{3}/2$ then we have an explicit
upper bound:
\[
2a_{1}+a_{k}+a_{k+2}-(2-\sqrt{3})a_{k+2}/2.
\]
This rules out PGST when $n=3k+2$ and $k$ is even.

\bigskip

\emph{Appendix 3 --} Through this work we have studied PGST between the
extremities of a chain. For practical purposes it may be useful to have PGST
between particles corresponding to internal nodes. We discuss an argument for
showing that if there is PGST between internal nodes then there is PGST
between the extremities. In doing so, we appeal to algebraic techniques from
quantum control theory of spin systems (Ref. [6]). Given a graph $X$ with set
of vertices $V$ and adjacency matrix $A$, let $z$ be the characteristic vector
of some set $S\subseteq V$. We define and denote by
\[
W_{z}=[z|Az|...|A^{n-1}z]
\]
an $n\times n$ matrix with entries in $\mathbb{Z}^{\geq0}$. The matrix $W_{z}%
$is called the \emph{walk matrix} of $X$ with respect to $S$. The pair
$\left(  X,z\right)  $ is said to be \emph{controllable} if the matrix $W_{z}$
is invertible (\emph{i.e.}, $\det(W_{z})\neq0$). The set-up is a
graph-theoretic analogue of the famous Kalman rank condition in control
theory:\ the matrices $e^{iAs}$ and $e^{izz^{T}t}$ ($s,t\in\mathbb{R}^{+}$)
from a controllable pair generate a dense subgroup of the unitary group
$U\left(  n\right)  $ ($n\geq2$).

A \emph{closed walk} is a sequence of vertices, with consecutive vertices
adjacent, that starts and ends at the same vertex. If we have PGST on $P_{n}$
from $k$ to $\ell$ then the vertices $k$ and $\ell$ are cospectral, so then
$k+\ell=n+1$ and the generating functions for closed walks at $k$ and at
$\ell$ are equal. Consequently, the distance from $k$ to an end vertex equals
the distance from $\ell$ to the antipodal end vertex.

By the $A$\emph{-module} generated by a vector, we mean the smallest
$A$-invariant subspace that contains the vector. If $|\ell\rangle$ lies in the
$A$-module generated by $|k\rangle$ and we have PGST from $k$ to $n+1-k$, then
we have PGST from $\ell$ to $n+1-\ell$.

Suppose%
\[
U(t)|k\rangle\approx\gamma|n+1-k\rangle.
\]
Because $|\ell\rangle$ lies in the $A$-module generated by $|k\rangle$, there
is a polynomial $f$ such that%
\[
|\ell\rangle=f(A)|k\rangle.
\]
Then
\[
U(t)|\ell\rangle=U(t)f(A)|k\rangle=f(A)U(t)|k\rangle\approx\gamma
f(A)e_{n+1-k}.
\]

If $\Phi$ is the `flip' automorphism on the path, then $\Phi$ commutes with
$A$ and so
\begin{align*}
f(A)|n+1-k\rangle &  =f(A)\Phi|k\rangle=\\
\Phi f(A)|k\rangle &  =\Phi|\ell\rangle=|n+1-\ell\rangle.
\end{align*}
If $n+1$ is a prime, then all vertices $i$ give a controllable pair
$(P_{n},i)$, and so PGST between any pair of vertices implies PGST between end
vertices. If $n+1=2p$ where $p$ is prime and $k\neq2,p$, then PGST from $k$ to
$n+1-k$ implies PGST between the end vertices. If $k=p$ then $k$ is the
central vertex and PGST cannot occur at $k$. We leave open the case
$n=2^{m}+1$.


\begin{thebibliography}{99}                                                                                               %


\bibitem {aks}M. Agrawal, N. Kayal, and N. Saxena, \emph{Ann. of Math.} (2)
\textbf{160} (2004), no. 2, 781-793.

\bibitem {m}I. Bloch, \emph{Nature} \textbf{453}, 1016 (2008); J. Majer,
\emph{et al.}, \emph{Nature} \textbf{449}, 443 (2007).

\bibitem {bo0}S. Bose, \emph{Phys. Rev. Lett.} \textbf{91}, 207901 (2003).

\bibitem {bo}S. Bose, \emph{Contemp. Phys.,} \textbf{Vol. 48} (1), pp. 13--30,
2007; A. Kay, \emph{Phys. Rev. A} \textbf{84}, 022337 (2011); V. M. Kendon and
C. Tamon, \emph{J. Comput. Theor. Nanosci.}, \textbf{8}(3):422-433, 2011.

\bibitem {bu0}D. Burgarth, S. Bose, \emph{New J. Phys.} \textbf{7} 135 (2005).

\bibitem {bu}D. Burgarth, D. D'Alessandro, L. Hogben, S. Severini, and M.
Young, arXiv:1111.1475v1 [quant-ph]; C. Godsil and S. Severini, \emph{Phys.
Rev. A} \textbf{81}, 052316 (2010).

\bibitem {c}M. Christandl, \emph{et al.}, \emph{Phys. Rev. Lett.} \textbf{92},
187902 (2004).

\bibitem {cj}J. H. Conway and A. J. Jones, \emph{Acta Arith.} \textbf{30}
(1976), 229-240; W. Watkins and J. Zeitlin, \emph{Amer. Math. Monthly},
\textbf{100}(5):471--474, May 1993.

\bibitem {di2000}D. DiVincenzo, \emph{et al.}, \emph{Nature} \textbf{408}, 339 (2000).

\bibitem {god1}C. Godsil, \emph{Discrete Math.} \textbf{312}(1): 129-147 (2012).

\bibitem {go}C. Godsil, arXiv:1011.0231v2 [math.CO]; N. Saxena, S. Severini,
and I. Shparlinski, \emph{Int. J. Quantum Inf.} \textbf{5}, 417 (2007).

\bibitem {hw}G. H. Hardy and E. M. Wright, \emph{An Introduction to the Theory
of Numbers} (Oxford University Press, Oxford, 2008), 6th ed.

\bibitem {her}I. N. Herstein, \emph{Topics in Algebra}, Wiley (New York), 1975.

\bibitem {kay}A. Kay, \emph{Int. J. Quantum Inf.} \textbf{8}, 641 (2010).

\bibitem {kea}J. P. Keating, \emph{et al.}, \emph{Phys. Rev. A} \textbf{76},
012315 (2007).

\bibitem {p03}N. Konno, \emph{Phys. Rev. E}, \textbf{72}, 026113 (2005); P. L.
Knight, E. Rold\'{a}n, and J. E. Sipe, \emph{Phys. Rev. A} \textbf{68},
020301(R) (2003).

\bibitem {l}E. Lieb, F. Wu, \emph{Phys. Rev. Lett.} \textbf{20} (1968), 1445-1448.

\bibitem {per}H. B. Perets, \emph{et al.}, \emph{Phys. Rev. Lett.}
\textbf{100}, 170506 (2008).

\bibitem {pl04}M. B. Plenio, J. Hartley, and J. Eisert, \emph{New J. Phys.} 6,
\textbf{36} (2004).

\bibitem {re08}P. Rebentrost, \emph{et al.}, \emph{New J. Phys.} \textbf{11},
033003 (2009).

\bibitem {sc09}H. Schmitz, \emph{et al.}, \emph{Phys. Rev. Lett.}
\textbf{103}, 090504 (2009).

\bibitem {sh2008}B. Shore, \emph{Acta Phys. Slovaca} \textbf{58}, 243 (2008).

\bibitem {ch}Y. Wang, F. Shuang, and H. Rabitz, \emph{Phys. Rev. A}
\textbf{84}, 012307 (2011).
\end{thebibliography}
\end{document}